\documentclass[print,structabstract]{aa}
\usepackage{graphicx}
\begin{document}
\title{The energy spectrum of anomalous X-ray pulsars and soft gamma-ray
repeaters}

\subtitle{}

\author{J. E. Tr\"{u}mper\inst{1}, A. Zezas\inst{2,3}, \"{U}. Ertan \inst{4},
\and
N. D. Kylafis\inst{2,3}}

\institute{Max-Planck-Institut f\"{u}r extraterrestrische Physik, 
Postfach 1312, 85741 Garching, Germany\\
\and
University of Crete, Physics Department, 71003 Heraklion, Crete, Greece\\
\and
Foundation for Research and Technology-Hellas, 71110 Heraklion, Crete, Greece\\
\and
Faculty of Engineering and Natural Sciences, Sabanc\i\ University, 
34956, Orhanl\i, Tuzla, \.Istanbul, Turkey
}

\date {Received ; accepted }


\abstract 
{Anomalous X-ray pulsars (AXPs) and soft gamma-ray repeaters (SGRs)
exhibit characteristic X-ray luminosities (both soft and hard) of around
$10^{35}$ erg s$^{-1}$ and characteristic power-law, hard X-ray
spectra extending to about 200 keV.  Two AXPs also exhibit pulsed 
radio emission.}
{Assuming that AXPs and SGRs accrete matter from a fallback disk, 
we attempt to explain both the soft and the hard X-ray emission as 
the result of the accretion process.  We also attempt to explain their radio
emission or the lack of it.}
{We test the hypothesis that the power-law, hard X-ray spectra are 
produced in the 
accretion flow mainly by bulk-motion Comptonization of soft photons emitted
at the neutron star surface.  
Fallback disk models invoke surface dipole magnetic
fields of $10^{12} - 10^{13}$ G, which is what we assume here.}
{Unlike normal X-ray pulsars, for which the
accretion rate is highly super-Eddington, the accretion
rate is approximately Eddington in AXPs and SGRs
and thus the bulk-motion Comptonization operates efficiently. 
As an illustrative example we reproduce both the hard and
the soft X-ray spectra of AXP 4U 0142+61 well using the XSPEC package compTB.}
{Our model seems to explain both the hard and the soft 
X-ray spectra of AXPs and SGRs, as well as their radio emission or the lack
of it, in a natural way.  
It might also explain the short bursts observed in these
sources.  On the other hand, it cannot explain the giant 
X-ray outbursts observed in SGRs, which may result from the conversion of 
magnetic energy in local multipole fields.}

   \titlerunning{High-Energy Emission from AXPs}

\authorrunning{Tr\"{u}mper et al. 2010}
\titlerunning{Energy Spectrum of AXPs and SGRs}

\maketitle


\section{Introduction}

Anomalous X-ray pulsars (AXPs) and soft gamma-ray repeaters (SGRs) constitute 
a special population of young neutron stars distinguished by  
much higher X-ray luminosities than their rotational powers and spin periods 
clustered in a narrow range (2 - 12 s). The AXPs and SGRs are now believed 
to belong to the same class of objects, since short bursts that were once 
believed to be a distinctive property of the SGRs were observed from some 
of the AXPs as well (Gavriil et al. 2002; Kaspi et al. 2003). They are all 
spinning down with spin-period derivatives in the $10^{-13} - 10^{-11}$ s 
s$^{-1}$ range (Woods \& Thompson 2006; Kaspi 2007; Mereghetti 2008 for 
recent reviews on AXPs and SGRs). 

In addition to these main properties, broad-band observations have revealed 
many other peculiarities of these sources, which provide constraints for the 
models. In particular, some of these sources are persistent with soft X-ray 
luminosities of $L_{\mathrm{x,soft}} \sim 10^{34} - 10^{36}$ erg s$^{-1}$, 
while others, discovered in recent years, are transients with 
$L_{\mathrm{x,soft}} \sim ~10^{33}$ erg s$^{-1}$ in quiescence (see Table 1 
of Mereghetti 2008). Both transient and persistent sources show occasional 
X-ray enhancements lasting from months to more than years 
and are correlated, in 
the long term, with infrared (IR) luminosities (Tam et al. 2004). The 
transient AXPs 
have been discovered in such X-ray outbursts (enhancements) when 
their $L_{\mathrm{x,soft}}$ levels were about two orders of magnitude higher 
than when in quiescence (Torii et al. 1998; Kouveliotou et al. 2003; 
Gotthelf et al. 2004; Ibrahim et al. 2004; Mereghetti et al. 2006, Israel 
et al. 2007; Muno et al. 2007). There are both similarities and systematic 
differences in the X-ray outburst light-curve morphologies of transient and 
persistent sources. Other constraints on models come from the properties of 
AXPs and SGRs in the optical to mid-IR bands and their relations with the 
X-ray luminosities.

These sources are widely believed to be magnetars mainly 
1) because they account for 
the energetics of the super-Eddington soft gamma-ray bursts and
2) because the $P$ and $\dot{P}$ measurements lead 
to large magnetic dipole strengths
based on the assumption that these sources spin down by magnetic dipole 
radiation. Magnetar models (Duncan \& Thompson 1992; 
Thompson \& Duncan 1993; 1995)
do explain the super-Eddington bursts of SGRs. However, the quantitative 
explanation of the persistent soft X-ray luminosity by magnetic field decay,
the optical and IR properties during persistent states, the X-ray and the 
accompanying IR enhancements, and the period clustering of AXP and SGRs 
all seem
to meet difficulties within the original frame of the magnetar model.

On the other hand, fallback disk models (Chatterjee et al. 2000; Alpar 2001)
can account for these observational facts. They have been developed in a 
series of connected, self-consistent ideas by means of detailed quantitative 
models (Ek\c{s}i \& Alpar 2003; Ertan \& Alpar 2003; Ertan \& Cheng 2004; 
Ertan et al. 2006; Ertan \& \c{C}al{\i}\c{s}kan 2006; Ertan et al. 2007; 
Ertan \& Erkut 2008, Ertan et al. 2009). These models do not explain the super-Eddington bursts.
Nevertheless, they are compatible with the presence of magnetar fields 
provided that these fields are in higher multipoles rather than in the dipole
component. In these models, the strength of the dipole magnetic field is found
to be less than $\sim 10^{13}$ G to account for the observations. The first 
direct observational support for the presence of fallback disks around these
systems came from SPITZER observations of AXP 4U 0142+61 in mid-IR bands by 
Wang et al. (2006). The same source has also been detected in optical and 
near-IR bands. Through model fits, Wang et al. (2006) show that the mid-IR
data can be reproduced by an irradiated disk model. These authors propose 
that the near-IR and the optical luminosities have a magnetic origin, while 
the mid-IR flux originates in a passive and irradiated disk. Later it was 
shown that all the data sets from optical to mid-IR can be accounted for by 
a single disk model that is active and irradiated by X-rays, which themselves
are produced by accretion of disk matter onto the neutron star (Ertan et al.
2007).  Recently, AXP 1E 2259+586 was also detected in the SPITZER bands.
The combined overall spectrum, including the earlier IR detections of this
source, is similar to that of AXP 4U 0142+61 (Kaplan et al. 2009).

In recent years, some of the AXPs, namely AXP 1E 1841--045 (Molkov et al. 2004;
Kuiper et al. 2004), AXP 1RXS J1708--40 (Revnivtsev et al. 2004; den Hartog et
al. 2008a) and AXP 4U 0142+61 (Kuiper et al. 2006; den Hartog et al. 2008b), 
have been found to emit pulsed hard X-rays up to 150 keV or more with isotropic
luminosities close to the soft X-ray luminosities 
(see den Hartog et al. 2008b and Mereghetti 2008). 

Detailed discussions of the magnetar models trying to explain the hard X-ray 
emission properties of AXPs can be found in (Heyl \& Hernquist 2005; 
Heyl 2007; Beloborodov \& Thompson 2007; Baring \& Harding 2007). The 
advantages and disadvantages of the different magnetar models are summarized 
in den Hartog et al. (2008b). 

The main aim of this paper is to discuss how the hard and the soft 
X-ray components can be produced in the framework of accretion from a 
fallback disk.  In doing so, we treat AXPs and SGRs as a class
with the following representative properties:

1) Luminosities are $\sim 10^{35}$ erg s$^{-1}$.

2) Soft ($< 10$ keV) and hard ($> 10$ keV) luminosities are similar.

3) Dipolar magnetic field strengths are $10^{12} - 10^{13}$ G.

4) Rotational periods are $\sim 5$ s.

We find that we can explain the observations very naturally by considering
the bulk-motion Comptonization (BMC) that takes place
in the accretion flow above the polar cap.  The seed photons for this
Comptonization are provided by the polar cap, while the observed soft
X-ray emission comes from both the polar cap and an extended region 
around it.

As an illustrative example, we fit the observed X-ray spectrum (both soft
and hard) of AXP 4U 0142+61 using the XSPEC package compTB (Farinelli
et al. 2008).  The fit is 
extremely good and the resulting parameters are very reasonable.

The BMC model is presented in \S\ 2.  
In \S\ 3 we present the X-ray data of AXP 4U 0142+61 and the fit to them 
using the XSPEC package compTB. 
In \S\ 4 we discuss our model, and in \S\ 5 we present the 
relation of AXPs and SGRs to other source classes.  Finally in \S\ 6 we 
give our conclusions.

\section{Bulk-motion Comptonization model}

\subsection{Soft X-ray emission}

The soft X-ray spectra of AXPs below 10 keV can be fitted by two-component 
models composed of two blackbody spectra or a blackbody plus a steep 
power-law spectrum (Mereghetti 2008). As pointed out by Gotthelf $\&$ Halpern
(2005), the double blackbodies are  more physically motivated. The hotter 
component may represent photospheric radiation from a small hot polar cap
(area $A_{\rm hot}$), while the cooler component is photospheric emission 
from a large fraction of the neutron star surface (area $A_{\rm cool}$). 
Actually, such a combination of spectra is also found for young energetic 
radio pulsars and for isolated neutron stars showing purely thermal emission.

For an accreting neutron star, whose magnetic axis is inclined with respect 
to the fallback disk, the emitting region $A_{\rm hot}$ at the base of
the accretion flow will not have radial symmetry around the magnetic pole, 
but will assume the bow-shaped configuration discussed by Basko \&
Sunyaev (1976) and confirmed by the MHD calculations of Romanova et al. 
(2004; see also Bachetti et al. 2009).  This region is heated by 
the infall of accreting matter and the produced soft X-ray photons either
escape unscattered, and are observed as such, or get upscattered in the 
accretion flow and produce the hard X-ray spectrum.

\subsection{Hard X-ray emission}

The equation describing upscattering of soft photons in a converging, 
ionized, fluid flow was 
first introduced and solved by Blandford \& Payne (1981a;b)
and Payne \& Blandford (1981).  Their 
Comptonization equation (bulk-motion Comptonization) was solved in the 
case of spherical accretion onto a neutron 
star by Mastichiadis \& Kylafis (1992), who assumed a reflective 
neutron-star surface.  This last work was generalized by Titarchuk et al.
(1996, 1997), who examined the general case of a partially reflecting 
inner boundary and also included a second order term in the flow 
velocity (see also Psaltis \& Lamb 1997; Psaltis 2001).

The idea in the bulk-motion Comptonization (BMC) model, as applied to
the AXP X-ray spectra, is the following.  Soft X-ray photons, emitted
by the polar cap region, where the accretion occurs, find themselves in
the accretion flow.  For accretion rates comparable to the Eddington
rate, the optical depth to electron scattering in the accretion flow
is close to unity or more. Thus, a fraction of the emitted soft photons
get trapped in the flow and after several, nearly head-on collisions
with the accreting electrons acquire significant amounts of energy.
As a result, a power-law, hard X-ray spectrum is produced.

Torrejon et al. (2004) have applied the BMC model to the wind-accreting, 
slowly spinning, neutron star 4U 2206+54, which has 
similar properties to our sources ($L_x \sim 10^{35}$ erg s$^{-1}$, 
$L_{\rm soft} \sim L_{\rm hard}$, $kT_{\rm bb} \sim 1.2$ keV, 
and a hard spectral tail extending up to $\buildrel > \over \sim 90$ keV).
The BMC model was  also used to fit hard X-ray tails observed in
neutron star sources by Paizis et al. (2006), and it was significantly 
improved by Farinelli et al. (2008).

\section{Application to AXP 4U 0142+61}

The AXP 4U 0142+61 is one of the brightest known AXPs. Therefore its 
spectrum can provide one of the most stringent tests of the applicability
of the BMC model to the X-ray spectra of AXPs.  It has been extensively 
observed with the {\textit{Chandra}}, XMM-{\textit{Newton}}, and INTEGRAL
X-ray missions. 

To constrain the low-energy (0.5 - 10.0 keV) spectrum of 4U 0142+61, 
we opted for the available {\textit{Chandra}} high-energy transmission 
grating (HETG) spectra. These data are least affected by pile-up, 
while they do cover the desired energy band. 
There are also {\textit{Chandra}} data obtained in 
continuous clocking mode (previously analyzed by Patel  et al. 2003); 
however, the calibration of this mode for spectral analysis is still 
uncertain, as is indicated by the differences between the published 
analysis of the HETG and the continuous clocking mode data (Juett et al. 
2006; Patel et al. 2003). 
For the high-energy spectrum of 4U 0142+61, we use the extensive ISGRI 
INTEGRAL data, which cover the 20 - 200 keV band.  The spectral analysis has 
been performed with the XSPEC v12.0 package (Arnaud 1996). All cited 
errors are at the 90\% confidence level for one interesting parameter 
unless otherwise specified.

\subsection{{\textit{Chandra}} data}
 
4U 0142+61 has been observed with the {\textit{Chandra}} High-Energy 
Transmission Grating for 25 ksec in May 2001 (OBSID 1018; PI C. Canizares).
We used the reduced spectra available from the {\textit{Chandra}} Grating 
Data Archive and Catalog (TGCat;    http://tgcat.mit.edu/). Since even the 
0th order spectrum is affected by pile up, we used the positive and negative
1st order of the Medium Energy Transmission Grating and the High Energy 
Transmission Grating. The individual spectra were binned in order to 
have at least 50 counts per bin. The four spectra were fitted 
simultaneously using the relevant response matrices (rmfs) and ancillary 
responce matrices (arfs) for each order.  

A fit of the HETG data with a two component blackbody and power-law model,
affected by photoelectric absorption by cold gas ({\sc{phabs}} model in 
XPSEC) gives a blackbody temperature of $0.42^{+0.03}_{-0.02}$ keV, a 
power-law slope of $\Gamma=3.66^{+0.25}_{-0.29}$, and an absorbing 
$\rm{H}{\textsc i}$ column density of  
$0.95^{+0.10}_{-0.12}\times10^{22}$ cm$^{-2}$.  These parameters are 
almost identical to the parameters derived by Juet et al. (2002) from the 
analysis of the same data ($kT=0.418 \pm 0.013$ keV; $\Gamma=3.3 \pm 0.4$,
$N_{\rm H} = (0.88 \pm 0.13) \times 10^{22}$ cm$^{-2}$). However, they are 
slightly different from the parameters derived by Patel et al. 
($kT=0.470 \pm 0.008$ keV; $\Gamma = 3.40 \pm 0.06$, 
$N_{\rm H}= (0.93 \pm 0.02)\times10^{22}$ cm$^{-2}$) from the spectral fits 
of the continuous clocking mode data. Given the uncertainties in the 
calibration of the latter mode, we consider that the HETG data provide a 
more accurate representation of the low-energy spectrum of   4U 0142+61.

\subsection{INTEGRAL ISGRI data}

4U 0142+61 has been extensively observed by INTEGRAL. We downloaded  
all public pointed observations from the INTEGRAL archive, which are longer 
than 3 ksec and the source falls within 10 degrees from the pointing 
direction. The data were analyzed with the OSA 
v8.0\footnote{http://www.isdc.unige.ch/integral/download/osa\_sw} data 
analysis software provided by the Integral Science Data Analysis Center. 
We used version 8.0.1 of the instrument characteristics database (which 
provides the latest calibration data) and version 30.0 of the reference 
catalog (which provides a list of known sources used for the source 
detection and spectral extraction processes). First, we produced an image 
of the  observed region in the 4 standard ISGRI bands (20 - 40 keV, 
40 - 60 keV, 80 - 100 keV, 100 - 200 keV) following the standard 
procedures described in the 
IBIS analysis user manual\footnote{http://isdcul3.unige.ch/Soft/download/osa/osa\_doc/osa\_doc-8.0/osa\_um\_ibis-7.0.pdf}.  
4U 0142+61 has been clearly detected in all four bands. Next we extracted 
a spectrum of  4U 0142+61, again following the standard procedures for IBIS 
data analysis. As recommended, we added 2\% systematic errors to the spectrum 
to account for calibration uncertainties. 

We fitted the spectrum with a power-law model, using the latest arf and 
rmf files. We find an energy slope of $\Gamma=0.94\pm0.10$, consistent with 
the slope of $\Gamma=1.05\pm0.11$ reported by Kuiper et al. (2006).  We 
also fitted the spectrum with older arf  files available in the instrument 
characteristics database  in order to assess the effect of different 
calibration data on the measured spectral parameters. We did not find any 
statistically significant difference between the estimated spectral 
parameters. 

\subsection{Joint {\textit{Chandra}} and INTEGRAL ISGRI fits}

In the previous paragraphs, we have shown that our analysis of the archival 
{\textit{Chandra}} and INTEGRAL ISGRI data of 4U 0142+61 nicely reproduces 
the canonical model consisting of an absorbed blackbody and power-law 
model in the soft X-ray band (0.5-10.0 keV) and a hard power-law above 
20.0 keV. Next we investigated whether the same spectrum can be reproduced with 
a bulk-motion Comptonization model. For this reason we used the model of 
Farinelli et al. (2008) provided as an external XSPEC model 
(model {\sc{compTB}}\footnote{http://heasarc.gsfc.nasa.gov/xanadu/xspec/models/comptb.html}). 
This model includes a self-consistent treatment of the seed blackbody 
spectrum and the thermal and/or bulk-motion Comptonization of its photons.
Therefore we did not include an ad-hoc blackbody component. The seed 
spectrum is described by a modified blackbody function $S(E) \propto 
E^{\gamma-1} / [exp(E/kT_{s}) -1]$, where $kT_{s}$ is the 
characteristic temperature of the blackbody, and $E^{\gamma}$ is a power-law
component that modifies the blackbody. For $\gamma=3$ this component 
simplifies to a pure blackbody. Since this model includes both bulk-motion 
and thermal Comptonization, an important parameter is the relative 
efficiency of the two components defined as 
$\delta = <E_{\rm bulk}>) / <E_{\rm th}>$. For $\delta=0$ we have a pure 
thermal Comptonization spectrum. Other parameters of this model are 
the energy index of the Comptonization spectrum ($\alpha$; for more details
see  Farinelli et al.  2008), the temperature of the Comptonizing electrons
($kT_{e}$), a factor describing the ratio between the observed Compton 
scattered spectrum and the observed seed blackbody spectrum ($A$), and 
the normalization of the seed photon spectrum ($C_{N}$). To model
photoelectric absorption by cold gas, we included the XSPEC {\sc{phabs}} 
model component. 

\begin{figure}
\centering
\includegraphics[angle=-90,width=9cm]{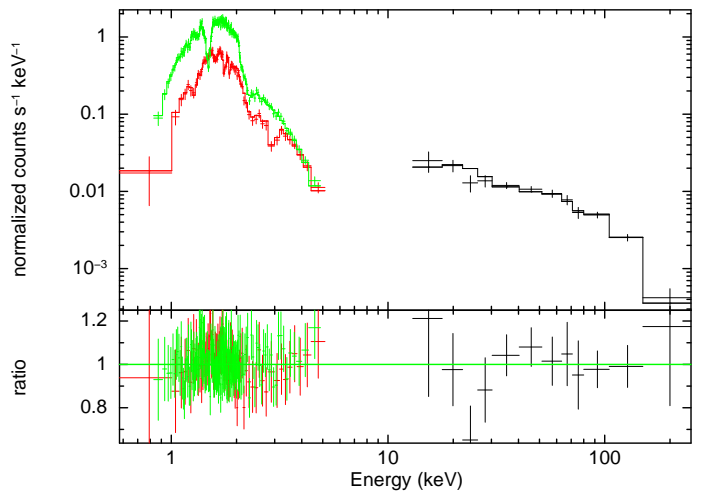}
\caption{Top panel: The {\textit{Chandra}} HEG and MEG (first positive 
order shown with red and green points respectively) and INTEGRAL ISGRI 
data (shown in black) along with the best-fit {\sc{comptb}} model. 
Bottom panel: Ratio between the data and the best-fit model.}
\label{Fig:Spec1}
\end{figure}

\begin{figure}
\centering
\includegraphics[angle=-90,width=9cm]{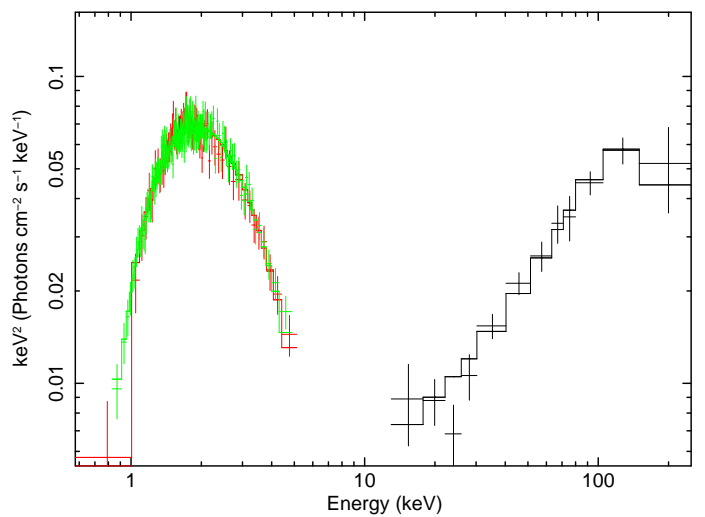}
\caption{The best-fit {\sc{comptb}} model, along with the data corrected for 
instrumental effects shown in $E^{2}f(E)$ space. }
\label{Fig:Spec2}
\end{figure}

The results of this fit are presented in Table \ref{Table:Spec}  and 
Figures \ref{Fig:Spec1} and \ref{Fig:Spec2}. The first figure shows the 
data and the best-fit model (top panel). It also shows the ratio of the data 
to the best-fit model (bottom panel).  Figure  \ref{Fig:Spec2} shows the 
model and the data corrected for instrumental effects (unfolded spectrum) in 
$E^{2}f(E)$ space. It is evident from this figure that the model gives a 
very good fit to 
the broad band spectrum of 4U 0142+61 ($\chi_{\nu}^{2}=166.7/248$). 
The temperature of the seed photons ($kT=0.72$ keV) is slightly higher than
the temperature estimated from the blackbody fit of the 
{\textit{Chandra}} data alone.  However, the parameter 
$\gamma=0.95$, which modifies the seed blackbody spectrum indicates that 
it is not a pure blackbody. Furthermore, we find that the 
bulk-motion Comptonization
dominates over thermal Comptonization ($\delta=2.32$) and that only 0.3\% of 
the observed photons have been Compton scattered [$\log(A)=-2.3$]. 
These results show that the spectrum of 4U 0142+61 can be reproduced 
equally well with a self-consistent BMC model with very reasonable parameters.  

\begin{table}
\caption{Fits of the ISGRI and {\textit{Chandra}} data with the 
{\sc{comptb}} model}
\label{Table:Spec}
\begin{tabular}{l l}
\hline\hline
Parameter & Value \\
\hline
Blackbody temperature ($kT_{s}$) & 0.72 keV  \\
Index of Seed photons ($\gamma$)       & 0.95  \\
Energy index of Compton spec. ($\alpha$) & 0.36 \\
Efficiency of bulk over thermal Compt. ($\delta$) & 2.32 \\
Electron Temperature ($kT_{e}$) & 24.6  keV \\
Ilumination factor (log$(A)$)  & -2.3 \\
Normalization ($C_{N}$)$\dagger$  & $1.18\times10^{-2}$ \\
 $\chi^{2}$/d.o.f.  & 166.74 / 248 \\
\hline
\end{tabular}

$\dagger$ Normalization of the seed-photon spectrum defined as 
$L_{39}/D_{10}^2$, where $L_{39}$ is its luminosity in units of 
$10^{39}\,\rm{erg\,s^{-1}}$, and $D_{10}$ is the distance to the 
source in units of 10 kpc
\end{table}

\section{Discussion}

The advantage of the bulk-motion Comptonization model is that it uses a 
proven concept, which has been applied to many cases including the slowly
pulsating source 4U 2206+54, which in turn shows a low-luminosity 
($10^{35}$ erg s$^{-1}$) and a hard X-ray tail like the AXPs (Torrejon et al. 
2004).  Another advantage is that the whole spectrum, from the soft thermal
component to the hard X-ray tail, is explained by a single, well-understood 
mechanism using as few parameters as possible.  The accreting matter both
heats the polar cap, which emits the seed photons for Comptonization,
and provides the high-energy electrons, which do the Comptonization.
Depending on detailed conditions, both the soft and the hard X-ray 
components have luminosities that are within the
same order of magnitude as observed. Since the soft and hard components
come from the same place at the neutron star surface, their maxima occur
at basically the same rotational phase.  This agrees well with the
observational fact that these sources show a very slow change in pulse
profile with energy (Kaspi 2007).  On the other hand, there are fine
structures in the energy-dependent pulse profiles, which can be explained 
by cyclotron resonance effects, as in the case of normal X-ray pulsars having
dipole magnetic fields of similar strengths to the ones postulated by our 
model for AXPs and SGRs.

The radio emission observed in two of the AXPs [1E 1547-54
(Camilo et al. 2007) and XTE J1810-197 (Halpern et al. 2005; Camilo
et al. 2006)] can also be explained by the BMC model.  Usually, it is thought 
that the radio emission of a neutron star is quenched by the accretion,
but there may be little or no overlap in our model between the circumpolar
zone of radio emission and the bow-shaped accretion zone
(Romanova et al. 2004).
Indeed, according to the canonical expression, the size of the radio-active 
polar cap is $r_{\rm pc} = 10^4 P^{-1/2} \sim 5 \times 10^3$ cm.  It seems
likely that the bow-shaped accretion region is farther from
the magnetic pole.  Therefore, both radio and X-ray emission may coexist.

A necessary requirement for a detectable radio flux is that the magnetic
dipole power $\dot E$ is strong enough.
Comparing $\dot E = (1/6 c^3) B^2 R^6 (2\pi/P)^4$,
where $B$ is the surface magnetic field, $R$ the radius of the neutron star, 
$P$ the rotational period and $c$ the speed of light, with the observed
radio luminosities of the two pulsars, we estimate the radio emission
efficiencies to be $\sim 10^{-4}$ and 
$\buildrel > \over \sim 0.03$ for 1E 1547-54 and XTE J1810-197, 
respectively, when assuming a polar magnetic field of $10^{13}$ G.
A lower magnetic field strength is also allowed in our model, because the
observed radio luminosity is not isotropic.
These efficiencies are quite large, but comparable to those of
the most efficient normal radio pulsars with periods $> 2$ s, taken from the
ATNF catalog http://www.atnf.csiro.au/research/pulsar/psrcat/
(Manchester et al. 2005).

The fact that most AXPs and SGRs have remained undetected in the radio band 
despite deep searches (Burgay et al. 2007) may be due to their long periods,
which may place them beyond the death line. Another reason could be that 
long-period radio pulsars have narrow beams with widths scaling with $P^{-1/3}$ 
(Lyne \& Manchester 1988). For a period of 6 s, the full beam is then only 
$\sim 7$ degrees. Therefore the probability of detecting the radio beam of 
an AXP is only a few percent.

\section{Relation to other source classes} 

The main point of this paper is to explain the hard X-ray spectra of AXPs
and SGRs as a result of accretion from a fallback disk. In addition,
we attribute the soft X-ray spectra of AXPs and SGRs
to thermal, photospheric radiation from the polar cap (hotter 
component) and from the bulk surface of the neutron star (cooler component).
This latter point leads us to the following thoughts:

The class of isolated neutron stars showing purely thermal emission that 
was discovered by ROSAT (XDINS, "Magnificent Seven") exhibits blackbody 
emission as well, though at lower temperatures (by a factor $\sim 10$) 
and lower luminosities (by $\sim 10^4$) compared with AXPs. 
Like the latter, the XDINs have long periods and strong magnetic fields 
($\ge 10^{13}$ G, Haberl 2007). Their period derivatives and proper motion data 
indicate ages of a few hundred thousand years, and they are not associated with 
supernova remnants. All these facts show that XDINs are older than AXPs, 
supporting the idea that some (or possibly all) XDINs have been 
AXPs before they exhausted their residual disk and have subsequently
cooled down. 

Another young neutron-star population, namely the central 
compact objects (CCOs, see review by Pavlov et al. 2004), 
are probably related to AXPs, SGRs, and XDINs as well.  
It was originally proposed by Alpar (2001) that similarities and 
differences in these neutron star classes could be 
explained if fallback disks with different properties are included in 
the initial parameters of the evolution models in addition to initial 
spin period and the magnetic moment of the neutron star.

\section{Conclusions}

Contrary to common belief, AXPs and SGRs may not be that different from normal 
accreting X-ray pulsars. The major differences between the two classes of 
X-ray sources are the extent of the hard X-ray spectrum and the X-ray 
luminosity, which is lower for AXPs/SGRs by two to three orders of magnitude.
The hard X-ray 
power-law spectral index is essentially the same for both classes, but the 
spectrum of normal X-ray pulsars extends to about 20 keV, while that of the 
anomalous X-ray pulsars extends to about 200 keV. In our model,
this difference is caused by the normalized (to the Eddington value) 
accretion rate being near 
unity in AXPs/SGRs, and bulk-motion Comptonization takes place and
produces the hard X-ray tail, which extends to $\sim 200$ keV.

Our model naturally explains the $\sim 100$ \% pulsations observed at 
$\sim 100$ keV, because they come from the accretion flow above the polar cap.
It also explains the similarity between AXPs/SGRs 
and XDINS, which has often been noted. It may be that XDINS are evolved 
AXPs with no remaining accretion disk.  The absence of cyclotron lines in the 
observed X-ray spectra is a natural consequence of our model.  For a dipole 
magnetic field of $\sim 10^{13}$ G, the electron cyclotron line would appear 
at $E > 100$ keV, where the photon statistics are not good enough.  On the 
other hand, a proton cyclotron line would appear at $E \sim 0.05$ keV, where 
interstellar absorption is huge. 

Our model could in principle explain the short bursts observed from AXPs 
and SGRs as the result of infalling lumps of matter with appropriate mass. 
We defer an analysis of this process to another paper.
On the other hand, our model does not explain the
giant bursts observed from AXPs and SGRs.  These may indeed require
very strong, magnetar-type magnetic fields. Our model does not exclude 
the existence of such super strong magnetic fields if they reside in multipole 
components.

Last but not least, we feel that the detection of a fallback disk around the 
neutron stars in AXP 4U 0142+61 and AXP 1E 2259+586 
gives strong observational support to our model.

\begin{acknowledgements}
\"{U}.E. and J.T. thank the Astrophysics Group of the University of Crete 
and FORTH for their hospitality. 
This research has been supported in part by EU Marie Curie project no. 39965, 
EU REGPOT project number 206469 and by EU FPG Marie Curie Transfer of Knowledge 
Project ASTRONS, MKTD-CT-2006-042722. \"{U}.E. acknowledges research support 
from T\"{U}B{\.I}TAK (The Scientific and Technical Research Council of Turkey) 
through grant 107T013 and support from the Sabanc\i\ University Astrophysics 
and Space Forum. 
\end{acknowledgements}

\end{document}